\documentclass[aps,preprint]{revtex4-1}
\usepackage{amssymb}
\usepackage{amsfonts}
\usepackage{amsmath}
\usepackage{graphicx}
\usepackage{epsfig}

\def\beq{\begin{equation}}
\def\eeq{\end{equation}}
\def\bea{\begin{eqnarray}}
\def\eea{\end{eqnarray}}

\begin{document}

\title{Dirac equation and the Melvin Metric}
\author{L. C. N. Santos and C. C. Barros Jr.}
\affiliation{Depto de F\'{\i}sica - CFM - Universidade Federal de Santa Catarina,\\
CP. 476 - CEP 88.040 - 900, Florian\'opolis - SC - Brazil}

\begin{abstract}
A relativistic wave equation for spin 1/2 particles in the Melvin
space-time, a space-time where the metric is determined by a magnetic field,
is obtained. The energy levels for these particles are obtained as functions
of the magnetic field and compared with the ones calculated with the Dirac
equation in the flat Minkowski space-time. The numeric values for some
magnetic fields of interest are shown. With these results, the effects of
very intense magnetic fields in the energy levels, as intense as the ones
expected to be produced in magnetars or in ultra-relativistic heavy-ion
collisions, are investigated.
\end{abstract}

\maketitle



\vspace{5mm}

\section{introduction}

In the past few years systems with extreme magnetic fields ($B>10^{15}G$)
have been proposed to exist. In the magnetar analysis, for example, magnetic
fields of the order of $10^{15}G$ at the surface \cite{mag1}, \cite{mag2}
and $10^{17}G$ at the center are expected to exist. In ultra-relativistic
heavy ion collisions, at $\sqrt{s}$= 200 GeV (at RHIC), the magnetic field
is expected to reach values as high as $10^{19}G$ \cite{hyion1}-\cite{hyion3}
and at LHC, at $\sqrt{s}$= 7 TeV, $10^{20}G$.

When studying particles inside this systems (eletrons inside stars, and so
on), the usual way is to solve the Dirac equation \cite{apl1}, \cite{apl2},
for example, and find the energy levels, and in general, the Minkowski
space-time is considered. A question that is quite intriguing, is if when
the magnetic field reach these values, some effect of the structure of the
space-time may be observed. This is the purpose of this work.

Including the magnetic field in the metric is not a trivial question but
some solutions exist, as the Melvin metric \cite{melv1}- \cite{melv4}, where
a magnetic universe, with a magnetic field in the $z$ direction, is
considered, or the Gutsunaev solution \cite{mank1}, \cite{mank2}, for a
magnetic dipole. Of course it is always possible to study an arbitrary shape
of the magnetic field and solve (at least numerically) the resulting
Einstein equations, but with the objective of finding analytical results,
and with this procedure, exploring this effect in a first approximation, we
will consider the Melvin metric and find the wave equations for Dirac
particles subjected to a magnetic field inside this metric.

So, this paper will show the following contents: In section II, a brief
review of the formulation of the Dirac equation in curved spaces will be
made. In section III the wave equation in the Melvin metric will be worked
out.  The results and conclusons of this work will be shown in section IV.


\section{Dirac equation in curved spaces}

In this section a brief review about the wave equation for spin 1/2
particles in curved spaces will be made. The basic formulation and the
equations that will be needed in the next sections will be shown.

A fundamental characteristic of the Dirac equation is its invariance under
Lorentz transformations, so, when studying Dirac particles in curved spaces,
it is interesting to preserve this aspect. One way to do that is by using
the tetrads $e_{\text{ \ }\mu}^{\left( a\right) }$ that may be defined in
order to satisfy the equation 
\begin{equation}
g_{\mu\nu}=e_{\text{ \ \ }\mu}^{\left( a\right) }e_{\text{ \ \ }\nu
}^{\left( b\right) }\eta_{\left( a\right) \left( b\right) },   \label{eq1}
\end{equation}

\noindent where $\eta_{\left( a\right) \left( b\right) }$ is the Minkowski
tensor, that represents a flat space-time, and $g_{\mu\nu}$ the metric
tensor related to a space-time that possesses an arbitrary geometry \cite%
{tet1}-\cite{tet3}.

Observing eq. $\left( \ref{eq1}\right) $, we may note that the tetrads may
be used in order to project vectors from the curved space-time in the flat
space-time with the expression 
\begin{equation}
V_{\mu}=e_{\text{ \ }\mu}^{\left( a\right) }V_{\left( a\right) } , 
\label{eq1b}
\end{equation}

\noindent that relates the form of a vector in different space-time
geometries.

As it was said before, the Dirac equation when written in the Minkowski
geometry posseses Lorentz symmetry, and we want that the equivalent
equation, written in a curved space-time possesses the same characteristic.
In fact, if we study the behavior of the elements of the Dirac equation
under transformations that preserve the desired symmetries,  we may
understand which are the quantities that need to be added or modified in the
equation.

We must note that a spinor transforms according to 
\begin{equation}
\psi\rightarrow\rho\left( \Lambda\right) \psi,   \label{eq2}
\end{equation}

\noindent where $\rho\left( \Lambda\right) =1+\frac{1}{2}i\varepsilon^{%
\left( a\right) \left( b\right) }\Sigma_{\left( a\right) \left( b\right) }$,
and $\Sigma_{\left( a\right) \left( b\right) }$ is the spinorial
representation of the generators of the Lorentz transformation, written in
terms of the $\gamma^{\left( c\right) }$ matrices, $\Sigma_{\left( a\right)
\left( b\right) }\equiv\frac{1}{4}i\left[ \gamma_{\left( a\right)
},\gamma_{\left( b\right) }\right] $ \cite{nakahara}. The idea is to
construct a covariant derivative $\nabla_{\left( a\right) }\psi$ that is
locally Lorentz invariant, that means that we need to impose the
transformation condition 
\begin{equation}
\nabla_{\left( a\right) }\psi\rightarrow\rho\Lambda_{\left( a\right) }^{%
\text{ }\left( b\right) }\nabla_{\left( b\right) }\psi.   \label{eq3}
\end{equation}

\noindent The usual way to obtain the form of the covariant derivative
operator of the spinor is by supposing the combination 
\begin{equation}
\nabla_{\left( a\right) }\psi=e_{\left( a\right) }^{\text{ }\mu}\left(
\partial_{\mu}+\Omega_{\mu}\right) \psi,\text{ }   \label{eq4}
\end{equation}

\noindent and considering the form of the operator $\Omega_{\mu}$
transformation 
\begin{equation}
\Omega_{\mu}\rightarrow\rho\Omega_{\mu}\rho^{-1}+\partial_{\mu}\rho%
\rho^{-1}.   \label{eq5}
\end{equation}

\noindent If we choose the combination of terms 
\begin{equation}
\Omega_{\mu}\equiv\frac{1}{2}i\Gamma_{\text{ }\mu}^{\left( a\right) \text{ }%
\left( b\right) }\Sigma_{\left( a\right) \left( b\right) }=\frac{1}{2}ie_{%
\text{ \ }\nu}^{\left( a\right) }\nabla_{\mu}e_{\text{ \ }}^{\left( b\right)
\nu}\Sigma_{\left( a\right) \left( b\right) },   \label{eq6}
\end{equation}

\noindent or in an equivalent form 
\begin{equation}
\Omega_{\mu}\equiv\frac{1}{2}i\Gamma_{\text{ }(a)\mu(b)}\Sigma^{(a)(b)},
\end{equation}

\noindent with the term $\Gamma_{\text{ }(a)\mu(b)}$ defined as 
\begin{equation}
\Gamma_{\text{ }(a)\mu(b)}=e_{(a)\nu}\left( \partial_{\mu}e_{(b)}^{\text{ \
\ }\nu}+\Gamma_{\text{ \ }\mu\lambda}^{\nu}e_{(b)}^{\text{ \ \ }\lambda
}\right) ,
\end{equation}

\noindent where $\Gamma_{\text{ \ }\mu\lambda}^{\nu}$ are the Christoffel
symbols, eq. $\left( \ref{eq5}\right) $ is satisfied. Consequentlly eq.  $%
\left(\ref{eq4}\right) $ satisfies the transformation condition demanded in $%
\left( \ref{eq3}\right) $ and then we get the final form of the covariant
derivative operator 
\begin{equation}
\nabla_{\left( c\right) }\psi=e_{\left( c\right) }^{\text{ }\mu}\left(
\partial_{\mu}+\frac{1}{2}ie_{\text{ \ }\nu}^{\left( a\right) }\nabla_{\mu
}e_{\text{ \ }}^{\left( b\right) \nu}\Sigma_{\left( a\right) \left( b\right)
}\right) \psi.   \label{eq7}
\end{equation}

\noindent We must remark that there are different definitions of covariant
derivatives of tensors in the literature when studying the covariant Dirac
equation, as for example in \cite{he1}-\cite{he3}.

Considering the Dirac equation in a flat space-time 
\begin{equation}
i\gamma^{\left( a\right) }\partial_{\left( a\right) }\psi-m\psi=0, 
\label{eq7b}
\end{equation}

\noindent and replacing the conventional derivative operator by the one
obtained in $\left( \ref{eq7}\right) $ we obtain the desired wave equation
for spin 1/2 particles in a curved space-time 
\begin{equation}
ie_{\left( a\right) }^{\text{ }\mu}\gamma^{\left( a\right) }\left(
\partial_{\mu}+\Omega_{\mu}\right) \psi-m\psi=0.   \label{eq8}
\end{equation}

\noindent If the particle is submited to an external electromagnetic field,
we may introduce this effect by a minimal coupling 
\begin{equation}
ie_{\left( a\right) }^{\text{ }\mu}\gamma^{\left( a\right) }\left(
\partial_{\mu}+\Omega_{\mu}+ieA_{\mu}\right) \psi-m\psi=0,   \label{eq8b}
\end{equation}

\noindent and the scalar action that leads to this equation is given by 
\begin{equation}
S=\int d^{4}x\sqrt{-g}\bar{\psi}\left[ i\gamma^{\left( c\right) }e_{\text{ \ 
}\left( c\right) }^{\mu}\left( \partial_{\mu}+\frac{1}{2}ie_{\text{ \ }%
\nu}^{\left( a\right) }\nabla_{\mu}e_{\text{ \ }}^{\left( b\right) \nu
}\Sigma_{\left( a\right) \left( b\right) }+ieA_{\mu}\right) -m\right] \psi. 
\label{eq9}
\end{equation}

It is usual to define the term $\gamma^{\mu}=e_{\left( a\right) }^{\text{ }%
\mu }\gamma^{\left( a\right) }$ as a Dirac matrix in a given curved
space-time and it is easy to verify that it satisfies the Clifford algebra 
\begin{equation}
\gamma^{\mu}\gamma^{\nu}+\gamma^{\nu}\gamma^{\mu}=\mathbf{1}g^{\mu\nu}. 
\label{eq10}
\end{equation}


\section{Wave equation in the Melvin Metric}

If now we want to obtain the wave equations shown in the last section, in a
space-time that has its structure determined by a magnetic field, an useful
idea wolud be to consider the Melvin metric \cite{melv1}-\cite{melv4}. The
Melvin metric is a solution of the Einstein-Maxwell equations of the general
relativity that represents a cylindrical magnetic universe. In his work \cite%
{melv1}, \cite{melv2}, Melvin considered a static magnetic field where its
lines lie in cylindrical surfaces perpendicular to the radial direction,
with intensity $\sim B_0$ in the vicinity of the symmetry axis and falls as
fast as $B_0/r^4$ far away from the axis. This solution has wide-ranging
applications in the literature, as, for example in the study of Kerr black
holes \cite{mbh}, or in cosmology, where the possibility of the interaction
of the magnetic field with the expansion of the universe may be considered,
as for example in \cite{melv5}, \cite{kast}.

The line element may be written in a cylindrically symmetric form (taking $c$%
=$G$=$\hbar$=1) 
\begin{equation}
ds^{2}=\Lambda\left( r\right) ^{2}dt^{2}-\Lambda\left( r\right)
^{2}dr^{2}-\Lambda\left( r\right) ^{-2}r^{2}d\phi^{2}-\Lambda\left( r\right)
^{2}dz^{2},  \label{eq11}
\end{equation}
where $\Lambda\left( r\right) =1+\frac{1}{4}B_{0}^{2}r^{2}$ and $B_{0\text{ }%
}$ is the magnetic field. This metric reflects the curvature of the
space-time, determined by the existence of a magnetic field in the $z$
direction. For this reason, a metric with axial symmetry is taken into
account.

In the limit of vanishing the magnetic field we have $\Lambda(r)=1$ and the
equation (\ref{eq11}) becomes 
\begin{equation}
ds^{2}=dt^{2}-dr^{2}-r^{2}d\phi^{2}-dz^{2},  \label{eq12}
\end{equation}

\noindent that is the flat Minkowski space-time written in cylindrical
coordinates. So, we may define 
\begin{equation}
\eta_{\left( a\right) \left( b\right) }=\left( 
\begin{array}{cccc}
1 & 0 & 0 & 0 \\ 
0 & -1 & 0 & 0 \\ 
0 & 0 & -1 & 0 \\ 
0 & 0 & 0 & -1%
\end{array}
\right)   \label{eq13}
\end{equation}

\noindent and choose a diagonal tetrad basis  $e_{\text{ }\mu}^{(a)}$ 
\begin{equation}
e_{\text{ }\mu}^{\left( a\right) } =\left( 
\begin{array}{cccc}
\Lambda\left( r\right) & 0 & 0 & 0 \\ 
0 & \Lambda\left( r\right) & 0 & 0 \\ 
0 & 0 & r\Lambda\left( r\right) ^{-1} & 0 \\ 
0 & 0 & 0 & \Lambda\left( r\right)%
\end{array}
\right) ,  \label{eq14a}
\end{equation}

\noindent for which the equation $\left( \ref{eq1}\right)$ is satisfied, and
then it is easy to determine its inverse form $e_{\left( a\right) }^{\text{
\ }\mu}$,

\begin{equation}
e_{\left( a\right) }^{\text{ \ }\mu} =\left( 
\begin{array}{cccc}
\Lambda\left( r\right) ^{-1} & 0 & 0 & 0 \\ 
0 & \Lambda\left( r\right) ^{-1} & 0 & 0 \\ 
0 & 0 & \Lambda\left( r\right) /r & 0 \\ 
0 & 0 & 0 & \Lambda\left( r\right) ^{-1}%
\end{array}
\right) .   \label{eq14b}
\end{equation}

We are interested in studying the effect of a magnetic field $B_{0}$, that
modifies the space-time geometry, as it is shown in the Melvin metric. If we
consider a Dirac particle inside this field, we may also investigate the
effect of the minimal coupling in the wave equation for this particle
considering the 4-potential $A_{\mu}$. In a flat space-time, a constant
magnetic field in the $z$ direction, $B_{0}$, that may be related to an
equivalent magnetic field in the Melvin metric by the equation (\ref{eq1b})
appears if the 4-potential has the only non-vanishing component given by 
\begin{equation}
A_{\phi}=-\frac{2}{B_0\Lambda(r)}.   \label{eq16}
\end{equation}

\noindent Observing that the term $\gamma^{\mu}\Omega_{\mu}$, in the curved
space wave equation, relative to the tetrad (\ref{eq14a}) is given by 
\begin{equation}
\gamma^{\mu}\Omega_{\mu}=\frac{\gamma^{\left( 1\right) }}{2\Lambda^{2}}\frac{%
\partial\Lambda}{\partial r}+\frac{\gamma^{\left( 1\right) }}{2r\Lambda}, 
\label{eq17}
\end{equation}

\noindent equation (\ref{eq8b}) becomes 
\begin{equation}
\left[ \frac{\gamma^{\left( 0\right) }}{\Lambda}\frac{\partial}{\partial t}+%
\frac{\gamma^{\left( 1\right) }}{\Lambda}\left( \frac{\partial}{\partial r}+%
\frac{1}{2\Lambda}\frac{\partial\Lambda}{\partial r}+\frac{1}{2r}\right) +%
\frac{\gamma^{\left( 2\right) }\Lambda}{r}\frac{\partial}{\partial\phi }+%
\frac{\gamma^{\left( 3\right) }}{\Lambda}\frac{\partial}{\partial z}+iM+iq%
\frac{\gamma^{(2)}\Lambda A_{\phi}}{r}\right] \psi=0.  \label{eq18}
\end{equation}


\noindent In the limit $\Lambda\rightarrow 1$, this equation reduces to the
usual Dirac equation for a free particle in a flat space-time  in a
cylindrical coordinate system. So, we may interpret equation $\left( \ref%
{eq18}\right) $ as a generalization of a Dirac equation for a particle
inside a magnetic field, an equation that has been extensively studied in
the literature, in many contexts, as for example \cite{magnetico1}-\cite%
{magnetico5}. We are interested in observing the corrections in the energy
spectrum, due to the alteration of the geometry of the space-time,
determined in eq. $\left( \ref{eq18}\right) .$

Making a transformation in equation (\ref{eq18}) 
\begin{equation}
\psi=\frac{1}{\sqrt{r\Lambda\left( r\right) }}\Phi,   \label{eq20}
\end{equation}

\noindent we obtain a simplified form 
\begin{equation}
\left[ \frac{\gamma^{\left( 0\right) }}{\Lambda}\frac{\partial}{\partial t}+%
\frac{\gamma^{\left( 1\right) }}{\Lambda}\frac{\partial}{\partial r}+\frac{%
\gamma^{\left( 2\right) }\Lambda}{r}\frac{\partial}{\partial\phi }+\frac{%
\gamma^{\left( 3\right) }}{\Lambda}\frac{\partial}{\partial z}+iM+iq\frac{%
\gamma^{(2)}\Lambda A_{\phi}}{r}\right] \psi=0,   \label{eq20b}
\end{equation}

\noindent where the usual representation for the gamma matrices is
considered 
\begin{align}
\gamma^{\left( 0\right) } & =\left( 
\begin{array}{cccc}
1 & 0 & 0 & 0 \\ 
0 & 1 & 0 & 0 \\ 
0 & 0 & -1 & 0 \\ 
0 & 0 & 0 & -1%
\end{array}
\right) ,\text{ }\gamma^{\left( 1\right) }=\left( 
\begin{array}{cccc}
0 & 0 & 0 & 1 \\ 
0 & 0 & 1 & 0 \\ 
0 & -1 & 0 & 0 \\ 
-1 & 0 & 0 & 0%
\end{array}
\right) ,  \notag \\
\gamma^{\left( 2\right) } & =\left( 
\begin{array}{cccc}
0 & 0 & 0 & -i \\ 
0 & 0 & i & 0 \\ 
0 & i & 0 & 0 \\ 
-i & 0 & 0 & 0%
\end{array}
\right) ,\text{ \ }\gamma^{\left( 3\right) }=\left( 
\begin{array}{cccc}
0 & 0 & 1 & 0 \\ 
0 & 0 & 0 & -1 \\ 
-1 & 0 & 0 & 0 \\ 
0 & 1 & 0 & 0%
\end{array}
\right) ,   \label{eq21}
\end{align}

\noindent that has no dependence in the $z$, $t$ and $\phi$ variables. We
will suppose a solution of the form 
\begin{equation}
\Phi=R\left( r\right) \exp[-i\sigma t+ip_{z}z+im\phi],   \label{eq22}
\end{equation}

\noindent where $\sigma$ is the energy of the system, that assumes positive
values for particles and negative values for antiparticles, $p_{z}$ is the
momentum, $m=\pm1,\pm2,\pm3,...,$ a quantum number, and $M$, the electron
mass. So, equation (\ref{eq18}) may be written in a explicit form 
\begin{align}
& i\left( +\frac{d}{dr}+\frac{m}{r}\Lambda^{2}-\frac{q}{r}%
A_{\phi}\Lambda^{2}\right) R_{4}+\left( +\sigma-M\Lambda\right)
R_{1}-p_{z}R_{3}=0, \\
& i\left( +\frac{d}{dr}-\frac{m}{r}\Lambda^{2}+\frac{q}{r}%
A_{\phi}\Lambda^{2}\right) R_{3}+\left( +\sigma-M\Lambda\right)
R_{2}+p_{z}R_{4}=0, \\
& i\left( -\frac{d}{dr}-\frac{m}{r}\Lambda^{2}+\frac{q}{r}%
A_{\phi}\Lambda^{2}\right) R_{2}+\left( -\sigma-M\Lambda\right)
R_{3}+p_{z}R_{1}=0, \\
& i\left( -\frac{d}{dr}+\frac{m}{r}\Lambda^{2}-\frac{q}{r}%
A_{\phi}\Lambda^{2}\right) R_{1}+\left( -\sigma-M\Lambda\right)
R_{4}-p_{z}R_{2}=0.
\end{align}

\noindent In a first approximation we will solve the equation neglecting
terms in higher orders of $r,$ and then, the system of equations may be
written as 
\begin{align}
- & i\left( +\frac{d}{dr}+\frac{m}{r}+\frac{mB_{0}^{2}}{2}r-\frac{2q}{B_{0}r}%
-\frac{qB_{0}}{2}r\right) R_{4}+\left( -\sigma+M\right) R_{1}+p_{z}R_{3}=0,
\\
- & i\left( +\frac{d}{dr}-\frac{m}{r}-\frac{mB_{0}^{2}}{2}r+\frac{2q}{B_{0}r}%
+\frac{qB_{0}}{2}r\right) R_{3}+\left( -\sigma+M\right) R_{2}-p_{z}R_{4}=0,
\\
- & i\left( +\frac{d}{dr}+\frac{m}{r}+\frac{mB_{0}^{2}}{2}r-\frac{2q}{B_{0}r}%
-\frac{qB_{0}}{2}r\right) R_{2}+\left( -\sigma-M\right) R_{3}+p_{z}R_{1}=0,
\\
- & i\left( +\frac{d}{dr}-\frac{m}{r}-\frac{mB_{0}^{2}}{2}r+\frac{2q}{B_{0}r}%
+\frac{qB_{0}}{2}r\right) R_{1}+\left( -\sigma-M\right) R_{4}-p_{z}R_{2}=0, 
\label{eq26}
\end{align}

\noindent This set of equations may be decoupled by multiplying the first
equation  by the expression 
\begin{equation}
-i\left( +\frac{d}{dr}-\frac{m}{r}-\frac{mB_{0}^{2}}{2}r+\frac{2q}{B_{0}r}+%
\frac{qB_{0}}{2}r\right)
\end{equation}

\noindent and then, using the second and the fourth to eliminate the terms
containing the spinors $R_{1}$, $R_{2}$ and $R_{3}$. The result is 
\begin{equation}
\left[ \frac{d^{2}}{dr^{2}}-\frac{m^{\prime }\left( m^{\prime }+1\right) }{%
r^{2}}+b\left( -m^{\prime }+\frac{1}{2}\right) -\frac{b^{2}r^{2}}{4}+\left(
\sigma ^{2}-M^{2}-p_{z}^{2}\right) \right] R_{4}\left( r\right) =0,
\end{equation}

\noindent where $b=mB_{0}^{2}-qB_{0}$ and $m^{\prime }=m-2q/B_{0}$. In a
similar way we derive equations for $R_{1},$ $R_{2},R_{3}$ that may be
resumed in the form 
\begin{equation}
\left[ \frac{d^{2}}{dr^{2}}-\frac{m^{\prime }\left( m^{\prime }\pm 1\right) 
}{r^{2}}+b\left( -m^{\prime }\pm \frac{1}{2}\right) -\frac{b^{2}r^{2}}{4}%
+\left( \sigma ^{2}-M^{2}-p_{z}^{2}\right) \right] R\left( r\right) =0,
\label{eq28}
\end{equation}

\noindent where the positive sign refers to $R_{2}\left( r\right) $ and $%
R_{4}\left( r\right) $, and the negative one to $R_{1}\left( r\right) $ and $%
R_{3}\left( r\right) .$

As we may observe, equation (\ref{eq28}) is similar to the Schr\"{o}dinger
equation, and it may be written as 
\begin{equation}
\frac{d^{2}R\left( r\right) }{dr^{2}}+\left( E-V_{ef}\right) R\left(
r\right) =0,   \label{eq29}
\end{equation}

\noindent where $E=\sigma^{2}-M^{2}-p_{z}^{2}$. The term $V_{ef}=\frac{%
m^\prime\left( m^\prime\pm1\right) }{r^{2}}-b\left( -m^\prime\pm\frac{1}{2}%
\right) +\frac{b^{2}r^{2}}{4}$  may be identified as an effective potential
and as we can see, the system has the form of an isotropic harmonic
oscillator.



In fact, the solution of the equation (\ref{eq29}) may be mapped into a
3-dimensional harmonic oscillator-like one in spherical coordinates. These
solutions are given in terms of the associate Laguerre polynomials 
\begin{equation*}
R\left( S\right) =\left( 
\begin{array}{c}
N_{1}S^{\left( 1/2-m^\prime/2\right) }e^{-\frac{S}{2}}L_{n}^{1/2-m^{\prime}}
\\ 
N_{2}S^{-m^{\prime}/2}e^{-\frac{S}{2}}L_{n}^{-1/2-m^{\prime}} \\ 
N_{3}S^{\left( 1/2-m^{\prime}/2\right) }e^{-\frac{S}{2}}L_{n}^{1/2-m^{\prime
}} \\ 
N_{4}S^{^{-m^{\prime}/2}}e^{-\frac{S}{2}}L_{n}^{-1/2-m^{\prime}}%
\end{array}
\right) , 
\end{equation*}

\noindent where $S=\frac{br^{2}}{2}$ and $N_i$ are normalization constants.
The energy spectrum, relative to this solution is 
\begin{equation}
E=\left( n+\frac{3}{2}\right) \frac{b}{2},\text{ }n=0,1,2,3,..., 
\label{eq30}
\end{equation}

\noindent and using the definition of $E$ given in eq. (\ref{eq29}), we
obtain the energy spectrum 
\begin{equation}
\sigma=\sqrt{M^{2}+p_{z}^{2}+2B_{0}q\left( n+s+\frac{1}{2}\right)
+2B_{0}^{2}m\left( n+s+\frac{1}{2}\right) },   \label{eq31}
\end{equation}

\noindent where $s=\pm$1/2 is the spin quantum number of the particle and $m$
= 1, 2, 3..., in order to keep the conservation of the parity. As it was
pointed before, equation (\ref{eq31}) is written with $c=G=\hbar=1$. By
making the conversion to the international system of units we have 
\begin{equation}
\sigma=\sqrt{M^{2}c^{4}+p_{z}^{2}c^{2}+2\hbar c^{2}B_{0}q\left( n+s+\frac {1%
}{2}\right) +4\pi\varepsilon_{0}\hbar^{2}GB_{0}^{2}m\left( n+s+\frac{1}{2}%
\right) }.   \label{eq32}
\end{equation}

\noindent where $\varepsilon_{0}$ is the vacuum permissivity constant. In a
system of units with $\hbar=c=1$ we have 
\begin{equation}
\sigma=\sqrt{M^{2}+p_{z}^{2}+2B_{0}q\left( n+s+\frac {1}{2}\right)
+GB_{0}^{2}m\left( n+s+\frac{1}{2}\right) },   \label{eq33}
\end{equation}

\noindent were, in the last term inside the square root, the Planck mass $m_p
$ appears scaling the magnetic field as $(B_0/m_p)^2$.


\section{Results}

Studying equation (\ref{eq31}) we may recover some literature results. When
the magnetic field and $p_{z}$ goes to zero, we obtain the expression for
the rest energy of the particle 
\begin{equation}
\sigma=Mc^{2}.   \label{eq33}
\end{equation}

Now, if the gravitational energy is neglected, the last term inside the
square root vanishes and we have 
\begin{equation}
\sigma=\sqrt{M^{2}c^{4}+p_{z}^{2}c^{2}+2\hbar c^{2}B_{0}q\left( n+s+\frac {1%
}{2}\right) },   \label{eq34}
\end{equation}

\noindent that corresponds to the usual spectrum obtained from a Dirac
equation in a flat space with a vector potential. This fact confirms our
initial considerations about eq. (\ref{eq18}), interpreting it as a
generalization of this equation. It is also possible to analyse the
situation where only the gravitational term is considered, 
\begin{equation*}
\sigma=\sqrt{M^{2}c^{4}+p_{z}^{2}c^{2}+4\pi\varepsilon_{0}%
\hbar^{2}GB_{0}^{2}m\left( n+s+\frac{1}{2}\right) }, 
\end{equation*}

\noindent and observing this equation it is easy to see that this term (when
the magnetic field appears) is a negligible correction for usual magnetic
fields (not so strong) for all energy levels.

But we are interested in studying the effect of extreme magnetic fields, as
intense as the ones found in magnetars, of the order of $B\sim
10^{15}-10^{16}G$ \cite{mag1}, \cite{mag2} or the fields expected to be
produced in heavy-ion collisions $B\sim 10^{19}G$ \cite{hyion1}-\cite{hyion3}%
. The results for some energy levels, obtained with the exact numerical
solution of  eq. (\ref{eq18}) (Melvin metric), considering some systems of
interest, are shown in Table 1. Observing the Table, we can see that the
effect of the magnetic field appears when $B$ increases and becomes
important when these fields are intense.

The same calculations have been performed in a flat space-time with a
minimal coupling. The results are shown in Table 2. As we can see, the
results are essentially the same, and, when comparing then with the ones
found in Table 1, we can only find deviations when considering fields of the
order of $B\sim 10^{19}G$ and quantum numbers as large as $n\sim 10^{30}$.
These results are shown in Fig. 1. Observing these results, the conclusion
that we may obtain, with a very good precision, is that the effect of the
magnetic field in the metric is very small and may be neglected.

As it was explained before, eq. (\ref{eq31}) is an approximation for small
values of $r$, but when comparing the results with the exact numeric
calculations that have been performed, up to the precision shown in Tables 1
and 2 the conclusions that we obtain are essentially the same, and then we
may use this equation as a very good approximation. The corrections from
terms with higher powers of $r$ appear beyond the precision shown in these
tables. This fact may be seen if one observes the approximations that have
been made when eq.(32)-(\ref{eq26}) have been obtained. Terms of the order $%
(B_0r)^{2n}$, where $n$ is an integer, appears. These terms increase with $%
r^2$, and reaches the maximum values with some $r_M$, that determines the
size of the considered system. For heavy-ion collisions, for example, $%
(B_0r_M)^{2}\sim 10^{-33}$, for pulsars, $10^{-11}$. The greater value
obtained was for magnetars, $10^{-7}$, and the effect of these corrections,
as it was said before, are always beyond the precision shown in the tables,
so, we may conclude that the accuracy of the approximation is good, even for
extremely large magnetic fields and levels.

\begin{figure}[hbtp]
\centerline{
\epsfxsize=120.mm
\epsffile{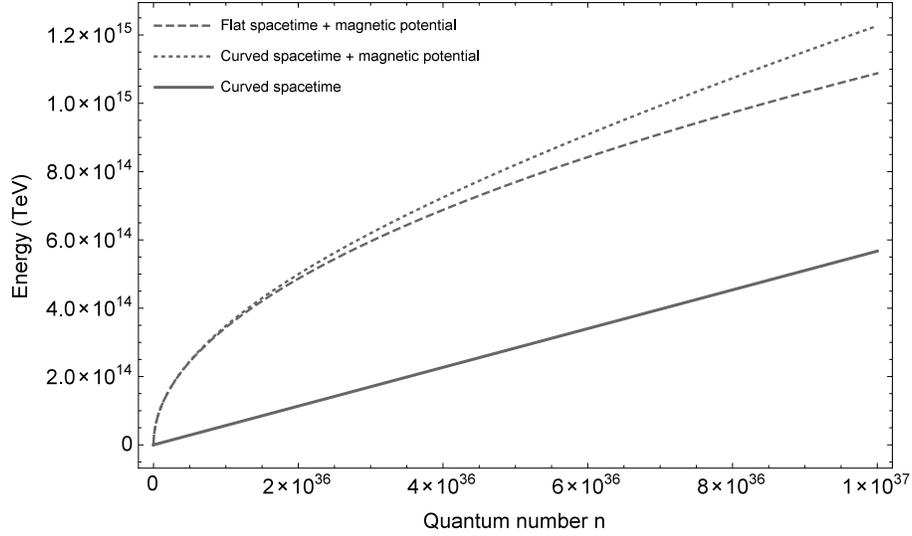}} 
\caption{Energy levels as functions of the quantum number $n$ considering
eq. (\ref{eq18}), (\ref{eq19}) and eq. (\ref{eq18}) without a 
minimal coupling (solid line).}
\end{figure}

\begin{figure}[hbtp]
\centerline{
\epsfxsize=120.mm
\epsffile{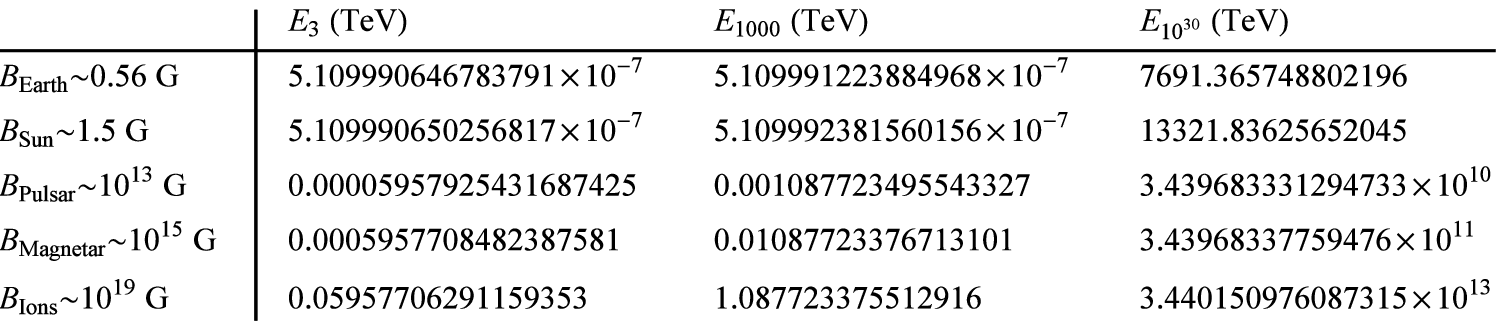}} 

{\small Table 1: Energy levels in the Melvin metric with a minimal coupling,
from the value of the magnetic field found in the Earth, up to the one
expected to be produced in ultra-relativistic heavy-ion collisions.}
\end{figure}

\begin{figure}[hbtp]
\centerline{
\epsfxsize=120.mm
\epsffile{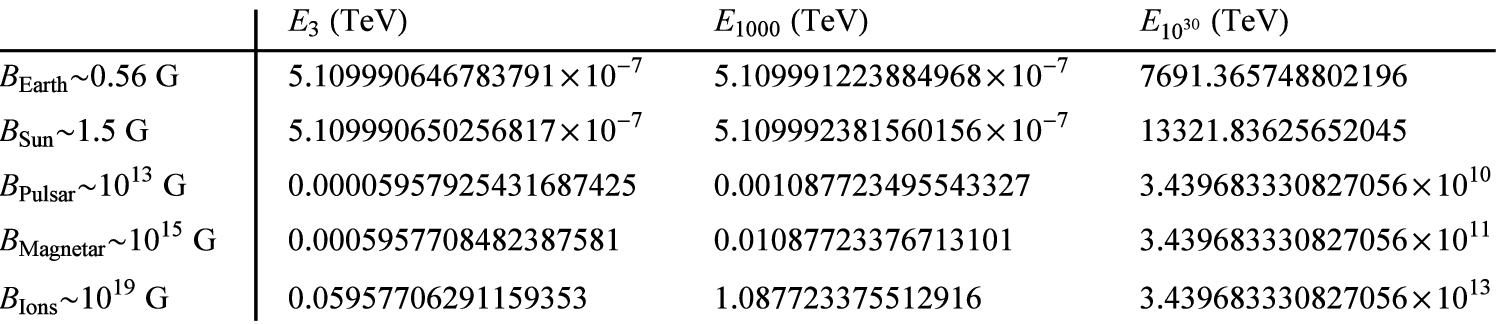}} 

{\small Table 2: Energy levels in a flat space-time.}
\end{figure}

One remark that must be made is that when considering a covariant Dirac
equation, a particle with gyromagnetic ratio $g=2$ is taken into account.
This is a good approximation for electrons inside stars, for example. But
when considering higher energy processes, as high-energy collisions,
deviations from this value are proposed to exist, and then, a way to study
this question is to consider extensions of the covariant Dirac equation, as
for example as it has been made in \cite{he3}. This kind of approach is left
for future works.

Another aspect that may be taken into account is the Melvin metric. It is
clear that many of the systems that have been studied does not have the
magnetic field in the form of the one that determines this metric, but in
some regions, with intense magnetic fields, at least as a first qualitative
approximation, these results must be correct, and we expect that more
careful calculations, with the metric determined by different shapes of the
magnetic  fields, confirm our results.

\vfill\eject


\end{document}